\documentclass[pre,groupedaddress,showkeys,showpacs,twocolumn,superscriptaddress]{revtex4}
\usepackage{amsmath}
\usepackage{graphics}
\usepackage{graphicx}
\usepackage{amsfonts}
\usepackage{siunitx}
\usepackage{amssymb}
\usepackage{xcolor}
\usepackage{dcolumn}
\usepackage{bm}
\begin{document}
\title{Swelling dynamics of surface-attached hydrogel thin films in vapor flows}
\author{Jessica Delavoipi\`ere}
\affiliation{Soft Matter Sciences and Engineering (SIMM), ESPCI Paris, PSL University, Sorbonne Universit\'e, CNRS, F-75005 Paris, France}
\affiliation{Saint-Gobain Research Paris, 39 quai Lucien Lefranc  93303 Aubervilliers Cedex, France}
\author{Bertrand Heurtefeu}
\affiliation{Saint-Gobain Research Paris, 39 quai Lucien Lefranc  93303 Aubervilliers Cedex, France}
\author{J\' er\' emie Teisseire}
\affiliation{Saint-Gobain Research Paris, 39 quai Lucien Lefranc  93303 Aubervilliers Cedex, France}
\author{Antoine Chateauminois}
\affiliation{Soft Matter Sciences and Engineering (SIMM), ESPCI Paris, PSL University, Sorbonne Universit\'e, CNRS, F-75005 Paris, France}
\author{Yvette Tran}
\affiliation{Soft Matter Sciences and Engineering (SIMM), ESPCI Paris, PSL University, Sorbonne Universit\'e, CNRS, F-75005 Paris, France}
\author{Marc Fermigier}
\affiliation{Physique et M\' ecanique des Milieux H\' et\' erog\`enes (PMMH), ESPCI Paris, PSL University, Sorbonne Universit\'e, CNRS, F-75005 Paris, France}
\author{Emilie Verneuil}
\affiliation{Soft Matter Sciences and Engineering (SIMM), ESPCI Paris, PSL University, Sorbonne Universit\'e, CNRS, F-75005 Paris, France}
\email{emilie.verneuil@espci.fr}
	\begin{abstract}
	Hydrogel coatings absorb water vapor - or other solvents - and, as such, are good candidates for antifog applications. In the present study, the transfer of vapor from the atmosphere to hydrogel thin films is measured in a situation where water vapor flows alongside the coating which is set to a temperature lower that the ambient temperature. The effect of the physico-chemistry of the hydrogel film on the swelling kinetics is particularly investigated. By using model thin films of surface-grafted polymer networks with controlled thickness, varied crosslinks density, and varied affinity for water, we were able to determine the effect of the film hygroscopy on the dynamics of swelling of the film. These experimental results are accounted for by a diffusion-advection model that is supplemented with a boundary condition at the hydrogel surface: we show that the latter can be determined from the equilibrium sorption isotherms of the polymer films. Altogether, this paper offers a predictive tool for the swelling kinetics of any hydrophilic hydrogel thin films.\\
	\end{abstract}
%
\maketitle
	%
	%
\section*{Introduction}
\label{sec:introduction}
Hydrogel coatings are suitable candidates for applications where both transparency and hydrophilicity are required. Indeed these hydrophilic polymer networks may absorb water by swelling to several times their dry thickness. As such, hydrogel coatings are possibly good candidates for anti-fog applications, as proposed by several teams \cite{grube_svenja_moisture-absorbing_2015,chevallier_characterization_2011,shibraen_anti-fogging_2016,park_sungjune_anti-fogging_2016,lee_zwitter-wettability_2013} who discussed their efficiency in terms of light transmission while submitted to warm humid air. Indeed, the coating should act as a reservoir for humidity and observation is made  \cite{shibraen_anti-fogging_2016,park_sungjune_anti-fogging_2016} that the higher the film thickness, the more water is absorbed and the later mist appears. Nevertheless, their efficiency to delay mist or fog formation in relation to their kinetics of swelling in humid atmosphere has not been studied yet. 

In this paper, we present experimental results and a model allowing for a complete description of the vapor transfer to hydrogel coatings, accounting for the physical-chemical characteristics of the polymer hydrogel. These results are then used to predict the kinetics of swelling of hydrogel coatings. 

The generic situation considered here is a coating onto a cold substrate in a vapor flow of higher temperature, for which the transfer of vapor to the coating results from both diffusive processes and the advective flow: it is therefore described by the diffusion-advection equation. If the vapor flow alongside the coating is fast enough, the vapor transverse transfer to the coating by diffusion is confined to a thin layer adjacent to the coating, the diffusion boundary layer. In this case, analytical solutions to the diffusion-advection problem from the literature can be used. Formally, the latter situation corresponds to high P\'eclet numbers Pe, which characterizes the ratio of advective transport by the mean flow to the diffusive transport. To solve the problem, the diffusion-advection equation must be supplemented by a boundary condition at the coating interface with the vapor. To do so, thin films of polymer networks made out of different hydrophilic polymers were prepared and characterized by their equilibrium sorption isotherms in water vapor. We will then determine the relevant boundary condition and show that an approximate analytical solution to the transport problem can be derived in the high P\'eclet number regime. We will extend our model to the low P\'eclet numbers thanks to a numerical resolution of the diffusion-advection problem.

As the hygroscopy of the polymer in the film is expected to change the kinetics of swelling, polymer networks with varied affinity for water are chosen. Besides, transfers within the coating may be affected by the glassy or rubbery state of the polymer: indeed, the diffusion of water within glassy polymers can be at least two orders of magnitude slower than in melts, while hygroscopy is also usually lower \cite{dupas_glass_2014}. Furthermore, some hydrophilic polymers that are glassy in the dry state become rubbery by absorbing water: this plasticization phenomenon may occur at room temperature. Hence, the glassy or melt state of the polymer, as well as the solvent induced glass transition - or plasticization - were tested in their possible effects on the water transfers to the coating. In order to study the effects of hygroscopy and glass transition, coatings made of 3 different hydrophilic polymers were prepared: poly(PEGMA) was chosen as a polymer of low hygroscopy and remaining molten at all water contents; poly(DMA) as a hygroscopic polymer undergoing glass transition and becoming rubbery for water contents above $\phi_{g,PDMA}$ ranging between $15\%$ and $20\%$ for temperatures between $0^\circ C$ and $25^\circ C$ ; poly(NIPAM) as a common thermo-responsive polymer which is hydrophilic at room temperature or below and which, similarly to poly(DMA), is glassy in the dry state at room temperature or below, and becomes rubbery for water content above $\phi_{g,PNIPAM}$ ranging between $23\%$ and $30\%$ for temperatures between $0^\circ C$ and $25^\circ C$.\\
The transport problem was addressed by working in a cell where the vapor characteristics (controlled laminar flow, temperature and concentration), the geometry and the coating temperature were controlled so that solutions to the advection-diffusion problems could be derived and compared to the measured coating swelling kinetics. Moreover, the experimental conditions were chosen to match those encountered in typical applications, making our study relevant to practical applications, as detailed in the last section of the paper.\\
Within this framework, during the most part of the coating swelling and for coatings thinner than a limit thickness, we were able to show that the advection-diffusive flux in the vapor controls the swelling kinetics while an homogeneous concentration profile is achieved across the coating thickness. In these conditions, we were able to provide a comprehensive description of the vapor absorption kinetics by hydrophilic polymer coatings.
%
%
\section*{Materials and method}
\label{sec:experimental}
\subsubsection*{Synthesis and characterization of surface-attached hydrogel coatings}

Poly(NIPAM), poly(DMA) and poly(PEGMA) hydrogel coatings were synthetized by crosslinking and grafting preformed polymer chains through thiol-ene click chemistry route as detailed elsewhere \cite{delavoipiere_poroelastic_2016,chollet_multiscale_2016,li_submicrometric_2015}. The coating is made by spin-coating of the ene-reactive polymer solution on thiol-functionalized silicon wafers with added dithiol crosslinkers. Homogeneous films with well-controlled thickness $e_{dry}$ are then obtained.

The thickness of the dry coatings were measured by ellipsometry for thin coatings ($e_{dry}<1.5\,\si{\mu m}$) and by profilometry for thicker coatings. It typically ranges between $1\,\si{\mu m}$ and $3\,\si{\mu m}$.  The swelling ratio was shown earlier \cite{li_submicrometric_2015} to be independent of the coating thickness for thicknesses higher than 200 nm, so that the swelling ratio $S$ was precisely measured from ellipsometry measurements carried out on thin coatings (dry thickness smaller than $1\,\si{\mu m}$) under dry nitrogen flow and under water: $S=\frac{e_{w}}{e_{dry}}$ with $e_w$ the coating thickness in water. For poly(DMA) coatings, the swelling ratio $S$ was varied within the range $[3\, ; \, 5]$. For poly(NIPAM) coatings, $S=7$ and for poly(PEGMA) coatings, $S=2.7$.
\begin{figure}
	\includegraphics[width=1 \columnwidth]{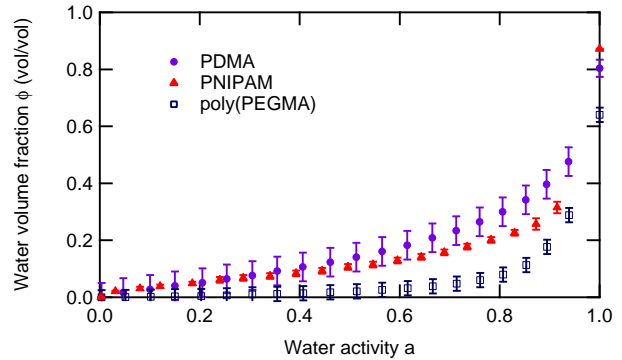}
	\caption{Sorption isotherms of poly(DMA) (S=5), poly(NIPAM) (S=7) and poly(PEGMA) (S=2.7) thin films at 20$^{\circ}$C. Water volume fraction absorbed by coatings made of the different polymers as a function of the water activity of the vapor at equilibrium, $a$, defined by Eq.1. The data at $a=1$ corresponds to the swelling ratio $S$ measured under water by ellipsometry.}
	\label{fig:dvs}
\end{figure}

The sorption isotherm of poly(NIPAM), poly(DMA) and poly(PEGMA) films were determined by classical Diffusion Vapor Sorption (Surface Measurement Systems) experiments. Prior to any measurement, the bare substrate is weighed in the DVS chamber, after what the thin film is prepared as detailed above. The sorption isotherm is then measured. Briefly, the coating is exposed to water vapor at a prescribed relative humidity which is increased step by step, and the corresponding water uptake is measured at equilibrium. Within experimental accuracy, no significant dependence of the sorption isotherms with temperature could be detected in a [12;25] $^{\circ}C$ temperature range. The results are plotted in Fig. \ref{fig:DVS}, where the water volume fraction $\phi$ in the coatings at equilibrium is plotted as a function of the water activity $a$, defined as the ratio of partial vapor pressure $p$ to the saturated water vapor pressure $p_{sat}(T)$:
\begin{equation}
a=\frac{p}{p_{sat}(T)}
\end{equation}
Water vapor is considered as an ideal gas for which $C_{sat}(T)=\frac{p_{sat}(T)M}{RT}$ where $C_{sat}(T)$ is the saturated water vapor concentration, $T$ the temperature in Kelvin, $M=18\,\si{g.mol^{-1}}$ the water molar mass, $R=8.31\,\si{J.mol^{-1}.K^{-1}}$ the ideal gas constant so that activity can be identified to relative humidity  $RH=\frac{C}{C_{sat}(T)}$
with $C$ the water vapor concentration. In this paper, the saturated water vapor pressure $p_{sat}(T)$ was calculated from the Rankine empirical formula, derived from Clapeyron model, with the latent heat of vaporization taken independent of temperature: 
\begin{equation}
p_{sat}(T)=p_{ref}e^{13.7-\frac{5120}{T}}
\end{equation}
with $p_{ref}=1\,\si{atm}$ the reference pressure.\\
The sorption isotherms in Fig.~\ref{fig:DVS} evidence the varied hygroscopy of the polymers used: of the three polymers, poly(DMA) is the most hygroscopic, and poly(PEGMA) the least.\\

\subsubsection*{Swelling kinetics measurement} 

\begin{figure*}[htb]
	\begin{centering}
		\includegraphics[width=2 \columnwidth]{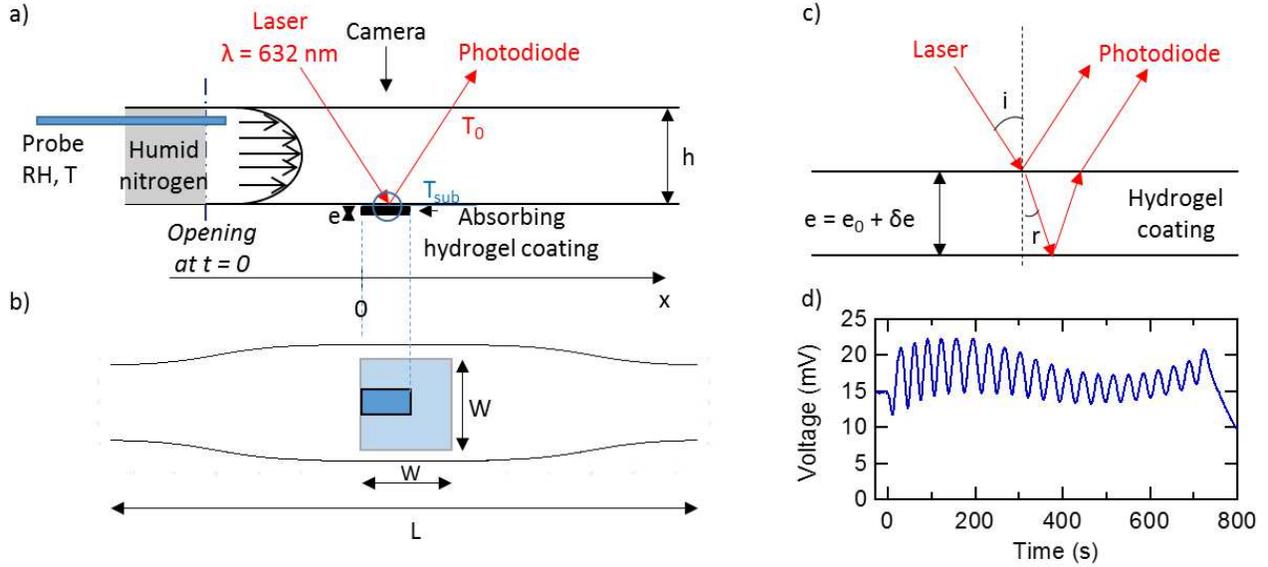}
		\caption{Schematic of the custom-built set-up. a) Side view: at $t < 0$, the absorbing coating is cooled down to $T_{sub}$ thanks to a Peltier module (not shown) and dried with dry nitrogen. At $t = 0$, humid nitrogen flow is sent in the Plexiglas chamber: the absorbing hydrogel coating swells and its thickness variation is deduced thanks to an interferometric measurement localized at $x=1\,\si{cm}$. Blue circle: zone zoomed in on c). The camera allows to observe the mist appearance. b) Top view (drawn to scale): the hydrogel coating (dark blue rectangle) is aligned with the leading edge of the Peltier module of dimensions W $\times$ W (light blue square). c) Principle of the thickness variation measurement with an interferometric set-up using a laser beam which reflects at the vapor/coating and the coating/substrate interfaces and is next collected by a photodiode. d) Photodiode voltage as a function of time during coating swelling.}
		\label{fig:montage_buee}
	\end{centering}
\end{figure*}

The water flux to hydrogel coatings was determined using a homemade setup (Fig. \ref{fig:montage_buee}) allowing a good control of the transport problem. A closed Plexiglas chamber was built up with length $L$ and width $W$ larger than its height $h$ to avoid side wall effects: $h=1\,\si{cm}$ ; $W=4\,\si{cm}$ and $L=28\,\si{cm}$. The chamber is equipped with a diverging  inlet and converging outlet on opposite sides (see Fig. \ref{fig:montage_buee}b)), to obtain a laminar flow of humid gas with controlled rate (measured with a flowmeter), and prescribed temperature and humidity (measured with a probe). Experiments with flow rates $Q$ of $0.1\,\si{L.min^{-1}}$ and $1\,\si{L.min^{-1}}$ corresponding to average velocities $U_0=\frac{Q}{Wh}$ of $4.2\, 10^{-3}\,\si{m.s^{-1}}$ and $4.2\, 10^{-2}\,\si{m.s^{-1}}$ were carried out. The corresponding Reynolds numbers $Re=\frac{U_0h}{\nu}$ with $\nu=1.5\,10^{-5}\,\si{m^2.s^{-1}}$ the kinematic viscosity of gas, are $3$ and $30$. We carefully paid attention to maintain a laminar flow at the highest Reynolds number by removing the temperature probe. In this case, the temperature is measured after the swelling experiments: it adds an uncertainty of $\pm 1^\circ C$ to the temperature in the chamber which is taken into account in the error bars. The humid gas is obtained by bubbling pure nitrogen in deionized water or in saturated salt solutions to adjust the relative humidity $RH$ which was varied between $20\%$ and $90\%$. The water vapor concentration of the incoming flow $C_0$ is defined as: $C_0=RH\times C_{sat}(T_0)$ with $T_0$ its temperature. Typically, $C_{sat}(T_0=25^\circ C)=22.4\,10^{-3}\,\si{kg.m^{-3}}$. The P\'eclet number $Pe$ characterizing the advection-diffusive problem then writes : 
\begin{equation}
Pe=\frac{U_0h}{D}
\label{eq:Pe_def}
\end{equation}
where $D=2.6\,10^{-5}\,\si{m^2.s^{-1}}$ is the water vapor diffusion coefficient in air.

In the bottom wall of the chamber, a Peltier module with dimensions $W\times W$ was inserted on which the samples were set. The leading edges of the sample and of the Peltier module were aligned. The Peltier module allows to create an homogeneous temperature field at the sample surface and to regulate the sample temperature to values lower than the ambient temperature, in the range 2 to 25$^\circ C$. Samples consist in silicon wafers with dimensions (typically $1\times 2\,\si{cm^2}$) smaller than the chamber sizes $W$ and $L$ on which model hydrogel coatings were prepared. Finally, the upper wall of the chamber is a glass window: it allows in a laser beam of wavelength $\lambda=632\,\si{nm}$, which reflects on the sample with an angle of incidence $i=50^\circ C$. The typical size of the laser spot on the sample is 400~$\mu$m. The reflected light is collected by a photodiode. Its intensity is modulated by the light interferences between the reflections on the coating/air and coating/substrate interfaces (Fig. \ref{fig:montage_buee} c)) by absorbing water, the coating swells and its thickness variation $\delta e$ is measured from the photodiode signal (Fig. \ref{fig:montage_buee}d)) using classical interferometry theory and assuming the gel optical index obeys a linear mixing law with water concentration (see SI for details). Observations from above using a CCD camera and a zoom allow for the detection of mist formation. Prior to any experiment, the sample is cooled down to temperature $T_{sub}$ and dry nitrogen flows in the cell. At initial time, humid nitrogen enters the chamber and the film thickness increase over time is measured, at a distance $x=1\,\si{cm}$ from the sample edge.
%
%
%
\section*{Experimental results}
\label{sec:discussion}
\begin{figure}
	\includegraphics[width=1 \columnwidth]{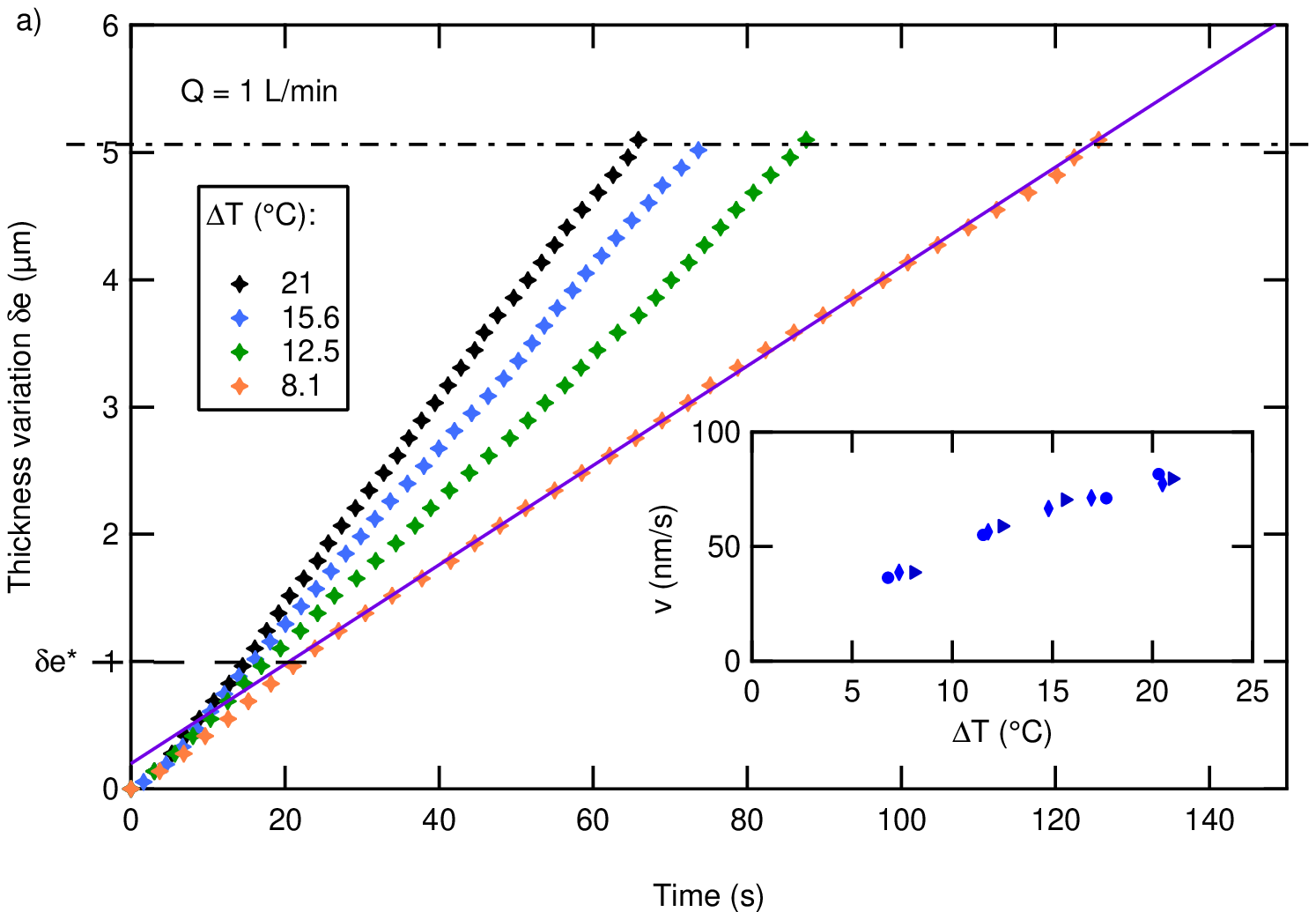}
	\includegraphics[width=1 \columnwidth]{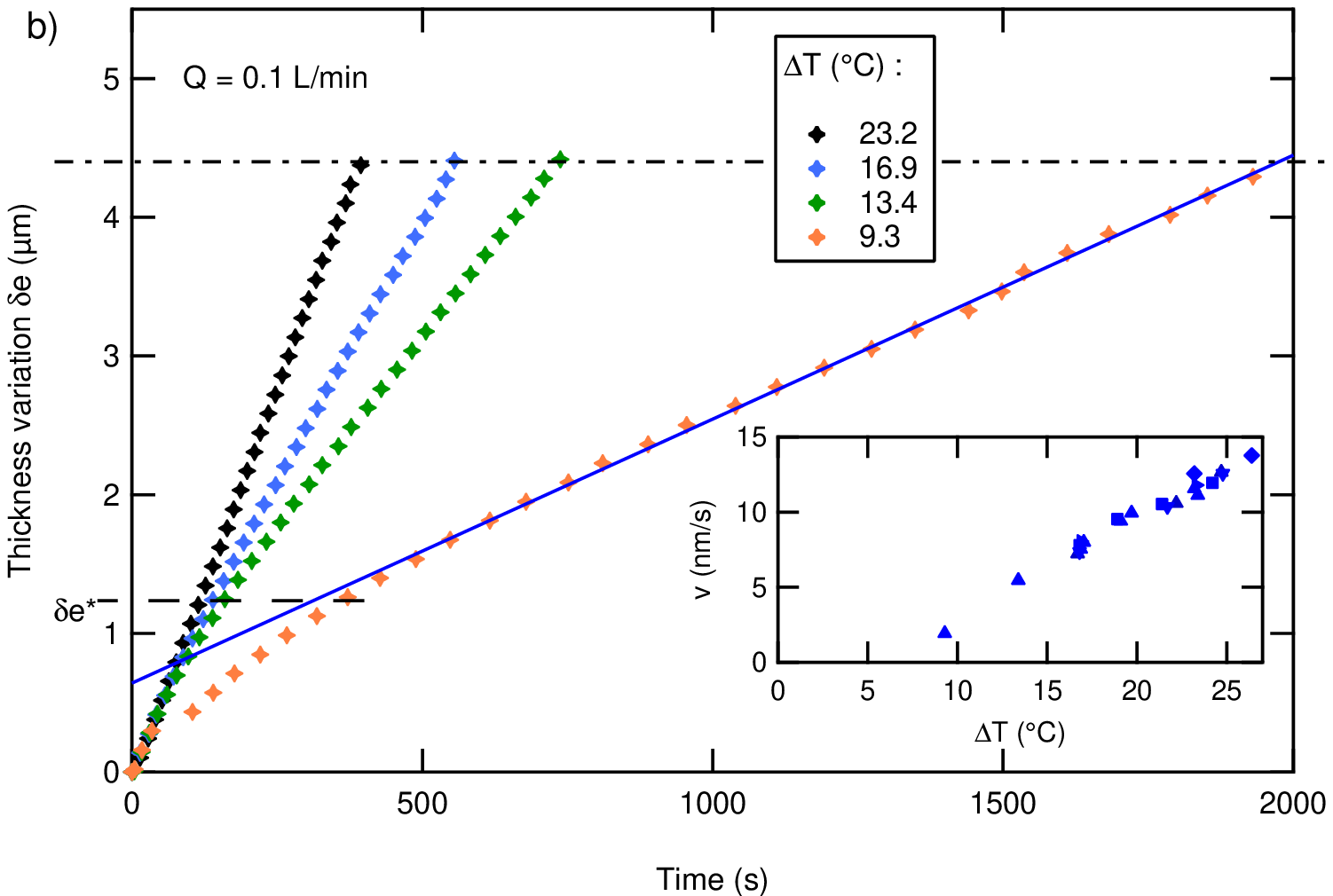}
	\caption{Swelling kinetics of a poly(DMA) film submitted to humid $N_2$ for various temperature differences $\Delta T=T_0-T_{sub}$. The dash-dotted line corresponds to the thickness increase for which mist appears. The solid line corresponds to a linear fit in the swelling linear regime. The dashed line marks the onset of the linear regime. a) $Q=1\,\si{L.min^{-1}}$, $Pe=16$, $RH = 88 \pm 1 \%$. Dry thickness: $e_{dry} = 1.5\,\si{\mu m}$, swelling ratio $S = 4.4$. b) $Q=0.1\,\si{L.min^{-1}}$, $Pe=1.6$, $RH = 60 \pm 1 \%$. $e_{dry} = 1.2\,\si{\mu m}$, $S = 5$. Insets: Swelling velocity determined in the linear regime as a function of the temperature difference $\Delta T$ for different dry thicknesses: \textcolor{black}{$\blacktriangledown$} 1000 nm (S=4.4) \textcolor{black}{$\blacksquare$} 1050nm (S=4) \textcolor{black}{$\blacktriangle$} 1200 nm (S=4 and 5) \textcolor{black}{$\blacktriangleright$} 1500 nm (S=4) \textcolor{black}{$\blacklozenge$} 2600 nm (S=4) \textcolor{black}{$\bullet$} 2970 nm (S=3)
	}
	\label{fig:delta_e_t_pdma_pe_16p2}
\end{figure}

First, due to the high aspect ratio of the films having a thickness $e$ much smaller than their lateral size, their swelling is one dimensional, and in the direction perpendicular to the plane of the substrate to which the polymer network is grafted.
Figure~\ref{fig:delta_e_t_pdma_pe_16p2} presents the increase in thickness $\delta e$ over time of a poly(DMA) coating for a given humidity of the incoming advective flow and various temperatures of the substrate $T_{sub}$ for $Pe=16$ (a) and $Pe=1.6$ (b). After a short transient regime, the thickness increases linearly over time. In the linear regime, the swelling velocity defined as $v=\frac{de}{dt}$ increases with the temperature difference $\Delta T$ between incoming air $T_0$ and substrate $T_{sub}$, as shown in inset in Fig.~\ref{fig:delta_e_t_pdma_pe_16p2} a). The coating thickness increment $\delta e$ can be related to the dry coating thickness $e_{dry}$, and to the averaged water volume fraction $\phi$. To do so, we assume a homogeneous water fraction along the coating thickness, a hypothesis that will be validated later. This gives: $\phi=\frac{\delta e}{e_{dry}+\delta e}$. For large enough temperature differences $\Delta T$ between the chamber temperature $T_0$ and the sample one $T_{sub}$, Fig.~\ref{fig:delta_e_t_pdma_pe_16p2} shows that the coating always swells to the same maximum thickness $e_{max}$ before mist appearance, as indicated by a dash-dotted line in Fig.~\ref{fig:delta_e_t_pdma_pe_16p2}. We check that this maximum thickness corresponds to the maximum swelling capacity $S$ of the hydrogel independently measured by ellipsometry on thin films. The same results were obtained for the three types of polymers. 

Besides, we observe that, for a given polymer, the linear regime for $\delta e(t)$ starts at the same value denoted $\delta e^*$, corresponding to the same average volume fraction $\phi^*$ for all experimental conditions. This result also holds for other humidities of the incoming flow tested. We determine $\phi^*$ for the three types of polymer: $\phi^*(poly(DMA))=40-45\%$ ; $\phi^*(poly(PEGMA))=20-30\%$ and $\phi^*(poly(NIPAM))=40-45\%$. For poly(DMA) and poly(NIPAM), $\phi^*$ is larger than the plasticization water fraction $\phi_g$.
\subsection*{Boundary condition at film interface}

From these observations, let us first validate the hypothesis of homogeneous water content along the coating thickness by evaluating the order of magnitude of the water concentration gradient along the coating thickness. 
To do so, let us write the equality of the water mass fluxes in the vapor $J_{vap}$ and across the coating $J_{polymer}$. The flux in the vapor corresponds to the diffusion across the boundary layer $\delta$ and scales as $DC_0/\delta$. As found in the literature \cite{bird_transport_2006}, at high P\'eclet numbers, the boundary layer scales with the channel height $h$ as $h/Pe^{1/3}$. With $Pe=16$ and $h=1\,\si{cm}$ here, we obtain: $\delta=4\,\si{mm}$. For a water volume fraction difference accross the film thickness $\Delta \phi$, the flux in the polymer scales as $\frac{\rho D_p\Delta\phi}{e}$ with $\rho=10^3\,\si{kg.m^{-3}}$ the water density and $D_p$ the water diffusion coefficient in the polymer coating: $D_p\approx10^{-11}\,\si{m^2.s^{-1}}$ for rubbery networks\cite{yoon_poroelastic_2010,salmon_humidity-insensitive_2017}. The equality of the water mass fluxes gives an estimate of the water volume fraction difference accross the film thickness: $\Delta \phi=\frac{e}{\delta}\frac{D}{\rho D_p}C_0$. At most, we find $\Delta\phi=0.1$ for a swollen film thickness $e=10\,\si{\mu m}$ and typical values $C_0=20\,10^{-3}\,\si{kg.m^{-3}}$, $\delta = 4\,\si{mm}$. Consequently, we will make the assumption that the water content in the coating is homogeneous along the thickness of rubbery films (poly(PEGMA) films at all water contents and poly(DMA) and poly(NIPAM) when molten, that is with $\phi_w>\phi_g$).

For glassy coatings, however, $D_p$ typically decreases to $10^{-14}\,\si{m^2.s^{-1}}$ and the water concentration gradient in the film thickness becomes significant. Consequently, at the beginning of the swelling of poly(DMA) and (poly)NIPAM coatings, diffusion within the coating is limiting. Poly(DMA) and poly(NIPAM) coatings undergo glass transition once a volume fraction of water $\phi_g$ of order 15$\%$ to 30$\%$ has been absorbed. This actually occurs early in the experiments, and always before the beginning of the linear regime since $\phi_g < \phi^*$. Meanwhile, in this early regime, the water vapor concentration profile in the chamber is in its transient: given the short duration of this complex transient regime compared to the total swelling time, it will not be further accounted for in this paper.\\

Let us now focus on the linear regime for the time variations of the coating thickness ($\delta e > \delta e^*$) for which water mass transfers are assumed to be limited by the flux in the vapor. 
This latter hypothesis is further confirmed by the fact that the measured swelling velocities are independent of the initial coating thickness (Fig.~\ref{fig:delta_e_t_PDMA_Pe_16p2}insets). Consequently, assumption can be made that the water fraction is homogeneous along the coating thickness holds at all times in the linear regime, and is equal to the average water volume fraction derived from theoretically measured $\delta e$. In addition, the evolution of the coating thickness can be ascribed to the water vapor flux $J_{vap}$ through:
\begin{equation}
J_{vap}=\rho\frac{de}{dt}=\rho v
\label{eq:Jvap_de_dt}
\end{equation}

The boundary condition in the vapor at the coating surface, namely the water vapor concentration at the surface $C_{sub}$, remains to be determined. If we report the experimental values of $\phi^*$ on the sorption isotherms (Fig. \ref{fig:DVS}), we note that it corresponds to activities ranging between 0.9 and 0.95 for all tested polymers. Hence, in the linear regime, the activity ranges between 0.9 and 1 and can therefore be considered as a constant denoted $a^*$. Meanwhile, the balance of water activities at the coating surface, in the polymer and in the vapor, writes:
\begin{equation}
a(\phi)=\frac{C_{sub}}{C_{sat}(T_{sub})}
\end{equation}

Considering a constant activity in the polymer enforces the water vapor concentration at the coating surface $C_{sub}$ to be constant all along the sample. In the following, we make the simple approximation that the water activity $a^*$ is equal to 1 in this regime, equivalently $C_{sub}\approx C_{sat}(T_{sub})$. We emphasize that in the narrow range of activity between $a^*$ and 1, ($a=1$ corresponds to the saturated gel), the water content of the gel varies strongly, as seen from Fig.~\ref{fig:DVS}.

Finally, hypothesis can be made that water condensation, if any, does not lead to a temperature increase through latent heat, so that the temperature at the film surface is constant. Also, the temperature dependence of the diffusion coefficient of water in air \cite{marrero_1972} will be neglected: indeed, a 10 $\%$ variation in $D$ is expected for $\Delta T=20^o$C. Any temperature effect will therefore be disregarded except for that on the saturation concentration of water vapor.\\

\subsection*{Transport problem}

In the following, we analyse the steady state transport from the advective diffusive vapor flow to the hydrogel coating.  The vapor volumetric concentration in the chamber is denoted $C$. The chamber width is larger than the sample width so that the problem reduces to a two dimensional problem: $C=C(x,y)$. As the sample is located in the middle of the chamber with $L/2>>h$, the gas flow is well-established above the sample. It can be described with a steady laminar parabolic velocity $\vec{u}=u(y)\vec{e_x}$ along the longitudinal direction x where:
\begin{equation}
u(y)=\frac{6U_0y}{h}(1-y/h)
\label{eq:v_parabol}
\end{equation} 
$U_0$ being the average velocity $U_0=\frac{Q}{Wh}$. While the downstream advection renews the vapor concentration to its initial concentration $C_0$, the diffusion in the cross-stream drives vapor to the coating where it is absorbed.
The steady state concentration of vapor is obtained by solving the advection-diffusion equation:
\begin{equation}
\vec{u}\cdot\vec{\nabla} C-D\Delta C=0
\label{eq:transport}
\end{equation}
with the boundary conditions:
\begin{equation}
C(y\rightarrow \infty)=C_0 \\
\label{eq:cond_lim1}
\end{equation} 
\begin{equation}
C(y=0)=C_{sub}
\label{eq:cond_lim2}
\end{equation} 

In the flow direction, the relative contribution of the advective and diffusive terms to the transport problem is characterized by the P\'eclet number as defined by the equation (\ref{eq:Pe_def}).

\begin{figure}
	\includegraphics[width=1 \columnwidth]{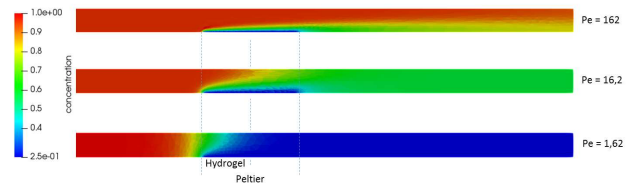}
	\caption{Concentration profiles in the vapor normalized by incoming concentration $C_0$ for Pe=162, 16.2, 1.62 from numerical simulations.}
	\label{fig:C_num}
\end{figure}

We first study the regime where $Pe>>1$. In this regime, we can make two hypotheses. In the flow direction, advection dominates over diffusion so that we neglect longitudinal diffusion. The concentration gradient develops in a boundary layer with a height $\delta$ small compared to the chamber height $h$. Within this depletion zone, the humid gas velocity is then linearized as:
\begin{equation}
u(y)=\frac{6U_0}{h} y
\label{eq:u_y_approx}
\end{equation}
Therefore, the advection diffusion problem reduces to:
\begin{equation}
\frac{6U_0}{h}y\frac{\partial C}{\partial x}=D\frac{\partial^2 C}{\partial y^2}
\end{equation}
Introducing the P\'eclet number, it becomes:
\begin{equation}
\frac{6Pe}{h^2}\frac{\partial C}{\partial x}=\frac{1}{y}\frac{\partial^2 C}{\partial y^2}
\label{eq:transport_simple}
\end{equation}
Autosimilar solutions of this classical Leveque problem \cite{bird_transport_2006} can be searched for in the form:
\begin{equation}
C=C_0f(\zeta)+C_{sub}
\label{eq:expr_conc}
\end{equation}
with the reduced variable $\zeta=\frac{yPe^{1/3}}{x^{1/3}h^{2/3}}$. Equation (\ref{eq:transport_simple}) becomes:
\begin{equation}
-2\zeta^2f'(\zeta)=f''(\zeta)
\end{equation}
By integrating this differential equation with the boundary conditions (\ref{eq:cond_lim1}) and (\ref{eq:cond_lim2}), we obtain:
\begin{equation}
C(\zeta)=C_{sub}+\left(C_0-C_{sub}\right)\frac{\int_{0}^{\zeta} e^{\frac{-2}{3}t^3}dt}{\int_{0}^{\infty} e^{\frac{-2}{3}t^3}dt}
\label{eq:sol_f}
\end{equation}
By noting the constant $A_2= \int_{l=0}^{\infty} e^{\frac{-2}{3}t^3}dt\sim 1.02$, the diffusive vapor flow toward the coating is then given by:
\begin{eqnarray}
J_{vap}(x)=D\left.\frac{\partial C}{\partial y}\right)_{y=0}=\frac{D(C_0-C_{sub})}{A_2}\left(\frac{U_0}{hxD}\right)^{1/3}\\
J_{vap}(x)=D\frac{C_0-C_{sub}}{\delta(x)}
\label{eq:flow_th}
\end{eqnarray}
with $\delta(x)=A_2\left(\frac{hxD}{U_0}\right)^{1/3}=A_2\left(\frac{xh^2}{Pe}\right)^{1/3}$ the thickness of the depletion layer. The diffusive vapor flow $J_{vap}$ is measured at the laser spot position ($x=1\,\si{cm}$). Our measurement averages the thickness variations over the laser spot size $\Delta x$ which is sufficently small compared to $x$ so that the corresponding variation $\Delta \delta$ of $\delta$ can be neglected ($\frac{\Delta \delta}{\delta}=\frac{1}{3}\frac{\Delta x}{x}\approx 10^{-2}$).
\\
For $Pe\approx1$ where the approximation $\delta<<h$ no longer holds, a solution of the advective diffusive problem is found by numerical resolution \cite{squires_making_2008} of the transport equation for the water vapor $\partial C/\partial t + \mathbf{u}.\nabla C = D \Delta C$ with the finite element code FreeFem++ \cite{hecht_new_2013}.
The velocity field is prescribed with the parabolic profile (equation (\ref{eq:v_parabol})). The initial condition is a constant concentration in the whole domain. At $t>0$, we impose a fixed normalized concentration equal to 1 at the inlet of the flow chamber, a concentration equal to $C_{sub}$ on the part of the bottom wall corresponding to the absorbing gel, and a zero normal flux at all other walls.  The concentration field is then iterated in time, until a steady-state situation is reached corresponding to the balance between the advection by the flow and molecular diffusion. At this point, the finite element mesh is refined according to the concentration gradient to resolve accurately the diffusion boundary layer on top of the adsorbing gel. From the concentration gradient at the bottom wall, we deduce the flux of vapor towards the gel. The concentration maps computed for different P\'eclet numbers are plotted in Fig.~\ref{fig:C_num}. Note that the numerical simulation ignores the limited absorption capability of the film.

%

\section*{Discussion}

\begin{figure}
	\includegraphics[width=1 \columnwidth]{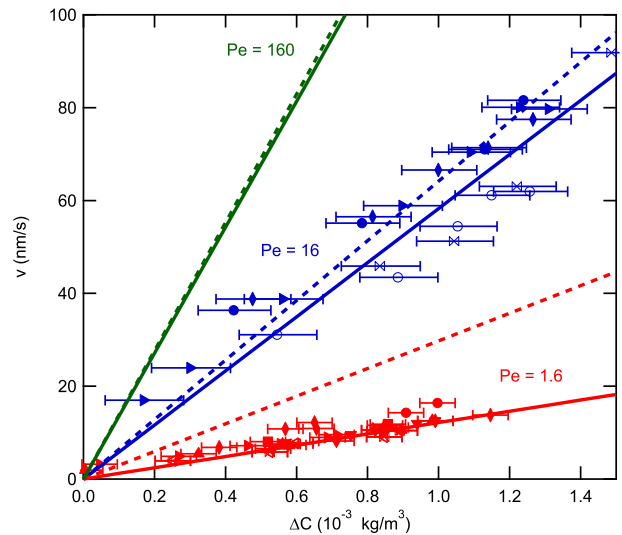}
	\caption{Experimental swelling velocities measured at $x=1\,\si{cm}$ as a function of $\Delta C = RH C_{sat}(T_{0})- C_{sat}(T_{sub})$ for experiments at $Pe = 16$ (blue markers) and $Pe = 1.6$ (red markers) with films of poly(PEGMA), PNIPAM, and PDMA of varied thickness and swelling ratio: poly(PEGMA): $\circ$ 910 nm, $\lhd$ 1500 nm ; PNIPAM: \textcolor{black}{$\Join$} 1450 nm ; PDMA: \textcolor{black}{$\blacktriangledown$} 1000 nm (S=4.4), \textcolor{black}{$\blacktriangle$} 1200 nm (S=4 and 5), \textcolor{black}{$\blacktriangleright$} 1500 nm (S=4), \textcolor{black}{$\blacklozenge$} 2600 nm (S=4), \textcolor{black}{$\bullet$} 2970 nm (S=3). Dotted lines and full lines represent respectively the analytical solutions (from eq (\ref{eq:flow_th})) and the numerical solutions for $Pe=1.6$ (red), $Pe=16$ (blue) and $Pe=160$ (green).}
	\label{fig:v_exp_deltac}
\end{figure}

Taking into account the previously determined boundary condition $C_{sub}=C_{sat}(T_{sub})$, Fig. \ref{fig:v_exp_deltac} represents the experimental swelling velocities $v$, or equivalently the vapor flux $J_{vap}/\rho$ through Eq.~\ref{eq:Jvap_de_dt}, as a function of $\Delta C = C_{0} - C_{sat}(T_{sub})$ for two series of experiments carried out at $Pe = 16$ and $Pe = 1.6$ respectively with different hydrogel films. The difference of concentration $\Delta C$ is varied by changing the temperature and/or the relative humidity RH of the humid nitrogen. At each P\'eclet number, we first find that all data collapse on the same curves within experimental accuracy. No effects of the thickness nor the swelling ratio $S$ is observed. This is consistent with the prediction in Eq.~\ref{eq:flow_th}. Furthermore, both approximated analytical solutions of the swelling velocity ($v_{anal}$) (from equation (\ref{eq:flow_th})) and  numerical solutions ($v_{simul}$) expressed by equations (\ref{eq:transport}) to (\ref{eq:cond_lim2}) are plotted for $Pe = 160, 16$ and 1.6. No experiments could be performed at $Pe=160$, but the simulated and analytical solutions at this  P\'eclet number were compared in order to test the approximated analytical solution. We see indeed that the analytical solution enables to describe the swelling at high P\'eclet numbers. While it is a very good approximation at $Pe=160$ ($v_{simul}/v_{anal}=0.98$), a $10\%$ difference is obtained for $Pe=16$. The analytical solution becomes out of range for $Pe=1.6$ ($v_{simul}/v_{anal}=0.41$) where longitudinal diffusion compares with advection. This is visible on the concentration maps in Fig.~\ref{fig:C_num} (Pe=1.62) where the thickness of the boundary layer extends over the flow cell height.

The experimental data with $Pe=1.6$ enable us to validate our model since the flow has likely less fluctuations at this low Reynolds number $Re=3$. The data at $Pe=16$ (open symbols) are in good agreement with the simulated prediction. The good agreement between the simulation and the experiments whatever the film nature, thickness and swelling ratio confirms that we can describe the swelling of an hydrogel coating cooled down and submitted to a humid air flow thanks to a diffusive-advective equation with the boundary condition: $C_{sub}=C_{sat}(T_{sub})$. 

These results can be used to predict the efficiency of these hydrogel coatings as anti-fog coatings. By choosing experimental conditions that are relevant to typical applications, our results can be readily transposed to design coatings (thickness, hygroscopy) with prescribed anti-fog capability. The delay to mist formation $T$ can be approximated by $T=\frac{\rho}{J_{vap}} \Delta e_{max}=\frac{\rho}{J_{vap}}e_{dry}(S-1)$ with $J_{vap}$ given by Eq.\ref{eq:flow_th} and $S$ and $e_{dry}$ are obtained from simple coating characterizations (DVS and profilometry).\\
For vertical plates (mirrors, windows,...) set at a temperature lower than humid atmosphere, our results can also be transposed by replacing the forced advective vapor flow with the thermal advective flow alongside the plate. Indeed, sizes and velocity conditions in these practical situations are such that the P\'eclet numbers are typically larger than 10. The vapor velocity gradient at the coating interface in Eq.~\ref{eq:u_y_approx} can be taken as the ratio between the advective velocity to the thermal boundary layer thickness. Our work thereby provides an efficient tool for the design of antifog absorbing coating made of hydrophilic polymers. \\
\section*{Supporting Information: Measurement of the hydrogel film thickness by interferometry}
The reflecting light resulting from the interferences between the reflections on the coating/air and coating/substrate interfaces depends on the phase shift $\varphi$ between the two reflected beams:
\begin{equation}
I=\frac{I_{min}+I_{max}}{2}(1+\Gamma \cos(\varphi))
\label{eq:intensity}
\end{equation} 
with $I_{max}$ et $I_{min}$ the maximum and minimum intensity, and $\Gamma$ the contrast ratio defined by:
\begin{equation}
\Gamma = \frac{I_{max}-I_{min}}{I_{max}+I_{min}}
\end{equation}
and $\varphi$ the phase shift:
\begin{equation}
\varphi=\pi+\frac{4\pi}{\lambda}ne\cos(r)
\end{equation}
with $e$ the thickness of the hydrated film and $n$ its refractive index. According to Snell and Descartes' law, the angle of the refracted beam $r$ in the film writes:
\begin{equation}
\cos(r(n))=\sqrt{1-\left(\frac{\sin(i)}{n}\right)^2}
\label{eq:cos_refrac}
\end{equation}
with $n$ the index of the hydrated polymer film. During the swelling, this index varies but the associated variation of $\cos(r)$ is negligible. We consider $cos(r)=cst=cos(r(n_{av}))$ with $n_{av}\approx 1.4$ the average refractive index. With the linear mixing law, the variation of the phase shift $\delta \varphi$ corresponds to the variation of the film thickness $\delta e$ so that:
\begin{equation}
\delta e \approx \frac{\lambda}{4\pi n_w\cos(r)}\delta \varphi
\label{eq:diff_marche}
\end{equation}
with $n_w$ the water refractive index. Between two consecutive extrema of the intensity (Eq. \ref{eq:intensity}) resulting from the interferences, the phase shift varies from $\pi$. Consequently, according to the equation (\ref{eq:diff_marche}), the corresponding thickness variation is:
\begin{equation}
\delta e_{min-max}=\frac{\lambda}{4 n_w\cos(r)}=141\,\si{nm}
\end{equation}
%


%
	%
\end{document}